\documentclass[a4paper]{article}
\usepackage{a4wide}
\usepackage{amsmath,amsfonts,amssymb,amsthm}

\numberwithin{equation}{section}

\newtheorem{theorem}{Theorem}[section]
\newtheorem{remark}[theorem]{Remark}

\newtheorem{proposition}[theorem]{Proposition}
\newtheorem{corollary}[theorem]{Corollary}
\newtheorem{lemma}[theorem]{Lemma}

\newcommand{\R}{{\mathbb R}}

\newcommand{\C}{{\mathbb C}}
\newcommand{\x}{{\bf x}}
\newcommand{\cst}{{\rm const}}
\newcommand{\A}{{\bf a}}

\newcommand{\y}{{\bf y}}
\newcommand{\z}{{\bf z}}
\begin{document}

\def\tr{{\rm Tr}}
\def\fl{{\rm fl}}
\def\Fl{{\rm Fl}}

\begin{center}

 \textbf{\Large Diamagnetic expansions for 
    perfect \\ quantum gases II: uniform bounds.}

\end{center}

\begin{center}
\textbf{Philippe Briet
\footnote{Universit\'e de Toulon et du Var, and Centre de Physique 
Th\'eorique-CNRS, Campus de Luminy, Case 907
                               13288 Marseille cedex 9, France; 
e-mail: briet@univ-tln.fr},
    Horia D. Cornean\footnote{Dept. of Mathematical Sciences,
    Aalborg
    University, Fredrik Bajers Vej 7G, 9220 Aalborg, Denmark;
e-mail:
    cornean@math.aau.dk}, and Delphine Louis.\footnote{
Centre de Physique Th\'eorique-CNRS, Campus de Luminy, Case 907
                               13288 Marseille cedex 9, France;
                               e-mail: 
louis@cpt.univ-mrs.fr}}
\end{center}

\begin{center}

 Revised version: December 17th, 2007

\end{center}

\vskip3mm\noindent {\bf Abstract: }%
{\small Consider a charged, perfect quantum gas, in the effective mass 
approximation, 
and in the grand-canonical ensemble. We 
prove in this paper that the generalized magnetic susceptibilities admit the 
thermodynamic limit for all admissible fugacities, uniformly on compacts 
included in the analyticity domain of the grand-canonical pressure. 

The problem and the proof strategy were outlined in \cite{3}. In 
\cite{4} we 
proved in detail the pointwise thermodynamic limit near $z=0$. 
The present  paper is the last one of this series, and 
contains the proof of the uniform bounds on 
compacts needed in order to apply Vitali's Convergence Theorem. }
\medskip


  
\section{Introduction and results.}%
The magnetic properties of a charged perfect quantum gas 
in the independent electron approximation and confined to a box $\Lambda$ 
have been extensively studied in 
the literature. One of the central problems has been to establish the 
thermodynamic limit for the magnetization and magnetic susceptibility, see 
e.g. \cite{1},  \cite{2}, \cite{3}, \cite{4} ,\cite{5}, \cite{6}, \cite{8}
and references therein. 

Briefly, the technical question is whether the thermodynamic limit ($\Lambda
\to \R^3$) commutes with the derivatives of the grand-canonical pressure 
with respect to the external constant magnetic field $B$.

A general way of proving this thermodynamic limit has been already announced 
in  \cite{3} and  \cite{6}, and the main ingredient consisted in applying 
the magnetic perturbation theory to a certain Gibbs semigroup. 
The strategy of the proof, which works  not only for the first 
two derivatives, but also for derivatives of all orders, 
were outlined in \cite{3}. In 
\cite{4} we proved in detail the pointwise thermodynamic limit near $z=0$. 
This  paper is the last one of the series, it 
contains the complete proof of the uniform bounds on 
compacts needed in order to apply the Vitali Convergence Theorem 
(see \cite{9}).

Now let us formulate the mathematical problem. The box which contains the 
quantum gas will be the cube $\Lambda\subset \R^3$ of side length $L>0$
centered at $0$. The constant magnetic field is ${\bf B}=(0,0,B)$, with 
$B\geq 0$, oriented parallel to the third component of the canonical 
basis in $\R^3$.

We associate to ${\bf B}$ the magnetic vector 
potential $B\A(\x)=\frac{B}{2}(-x_2,x_1,0)$ and the cyclotronic frequency
$\omega= \frac{e}{c}B $. In the rest of the paper, $\omega$ will  be a 
 real parameter. The one particle Hamiltonian we consider is the
self-adjoint operator densely defined in $L^2(\Lambda)$:
\begin{equation}\label{hamiunu}
H_L(\omega):=\frac{1}{2}\,(-i\nabla-\omega \A)^2,
\end{equation}
corresponding to Dirichlet boundary conditions. 

One denotes by $B_1(L^2(\Lambda))$ the Banach space of trace class operators. 
At $\omega\geq 0$ fixed, the magnetic Schr\"odinger operator $H_L(\omega)$ 
generates a Gibbs semigroup
$\{W_L(\beta,\omega)\}_{\beta\geq 0}$ where:
\begin{align}\label{defsemigg}
W_L(\beta,\omega):=e^{-\beta H_L(\omega)},\quad 
\|W_L(\beta,\omega)\|_{B_1}
\leq \frac{L^3}{(2\pi\beta)^{\frac{3}{2}}},\quad \beta >0.
\end{align}

Let us now introduce the grand-canonical formalism. Let $\beta=1/(kT) >0$ 
be the inverse temperature, $\mu\in\R $ the chemical potential and 
$z=e^{\beta\mu}$ 
the fugacity. Let $K$ be a compact included in the domain 
$D_+:=\C\backslash\,[e^{\frac{\beta\omega}{2}},\infty[$ for the Bose statistics
and $D_-:=\C\backslash\,]-\infty,-e^{\frac{\beta\omega}{2}}]$ for the 
Fermi case. 

Fix $\omega_0>0$ and  a
compact real interval $\Omega$ containing $\omega_0$. For a fixed $\beta >0$, 
one can find a simple, positively oriented, closed contour 
$\mathcal{C}_K\subset D_\pm$ whose interior 
does not contain $1$ in the Bose case or 
$-1$ in the Fermi case, and such that 
\begin{equation}\label{uniffbou}
\sup_{L>1}\sup_{\omega\in\Omega}\sup_{\xi\in \mathcal{C}_K }\sup_{z\in K}
\left \Vert [\xi-z
  W_L(\beta,\omega)]^{-1} \right \Vert =M<\infty.
\end{equation}
Details may be found in \cite{5} for the Bose case, 
but the main idea is that the spectrum of 
$W_L(\beta,\omega)$ is always contained in the interval 
$[0,e^{-\beta\omega/2}]$ 
and one can apply the spectral theorem. 

We then can express the grand canonical
pressure  at $\omega_0$ as follows (see e.g. \cite{5} for the Bose case): 
\begin{align}\label{Plambda}
P_L(\beta,z,\omega_0)=\frac{-\epsilon}{2i\pi\beta L^3}\int_{\mathcal{C}_K}
d\xi\,\frac{\ln(1-\epsilon\xi)}{\xi}{\rm Tr}\left [(\xi-z
  W_L(\beta,\omega_0))^{-1} z W_L(\beta,\omega_0)\right].
\end{align}
where $\epsilon=1$ for the Bose gas, and $\epsilon=-1$ for the Fermi
gas. 

For  $ \omega \in \R$, $\xi\in \mathcal{C}_K$ and $z\in K$, introduce the operator:
\begin{equation}\label{gebetatauom}
 g_L(\beta,z,\xi,\omega):=[\xi-z
  W_L(\beta,\omega)]^{-1} z W_L(\beta,\omega).
\end{equation}
This is a trace class operator which obeys (use \eqref{uniffbou} and 
\eqref{defsemigg}): 
\begin{eqnarray}\label{normeI1gL}
\|g_L(\beta,\omega,\xi,\omega)\|_{B_1}\leq 
  (\sup_{z\in K}|z|)\,\frac{L^3 M}{(2\pi\beta)^{\frac{3}{2}}}.
\end{eqnarray}
uniformly in $\omega \in \Omega$. Because $\omega \to W_L(\beta, \omega)$ is a ${B_1}$-entire operator valued 
function in $\omega$
(this result was first obtained in \cite{1} and then refined in 
\cite{4}), then using \eqref{uniffbou} one  easily shows that the map
$$]0,\infty[\ni\omega\mapsto {\rm Tr}\;g_L(\beta,z,\xi,\omega)\in\C $$
is smooth, with derivatives which are uniformly bounded in $\xi$ and $z$. 
Thus for every $N\geq 1$ and $z\in K$ 
we can define the generalized susceptibilities at $\omega_0$ by
\begin{align}\label{deffkon}
\chi_L^N(\beta,z,\omega_0)&:=\frac{\partial^N
  P_L}{\partial\omega^N}(\beta,z,\omega_0)\\
& =\frac{-\epsilon}{2i\pi\beta L^3}\int_{\mathcal{C}_K}
d\xi\,\frac{\ln(1-\epsilon\xi)}{\xi}\;\frac{\partial^N{\rm Tr}\;g_L
}{\partial\omega^N}(\beta,z,\xi,\omega_0).
\nonumber 
\end{align}
From this discussion one can  see that  the pressure as well  $\chi_L^N(\beta,\cdot,\omega_0), N\geq 1$ are  analytic functions 
on $D_+$ (or $D_-$). 

Now let us describe the case when $L=\infty$. The thermodynamic limit 
of the pressure exists and is uniform on compacts like $K$. If 
$\omega >0$, its actual value is (see \cite{2} ):
\begin{equation}\label{corc1000}
P_{\infty}(\beta,z,\omega_0):= \omega_0 \frac{1}{(2\pi\beta)^{3/2}}
\sum_{k=0}^\infty f_{3/2}^{\epsilon}\left (z e^{-(k+1/2)\omega_0\beta}\right ),
\end{equation}
where $f_\sigma^{\epsilon}(\zeta)$ are the usual Bose (or Fermi) functions for 
$\epsilon=1$ (or $\epsilon=-1$):
\begin{equation}\label{gbosonic}
f_\sigma^{\epsilon}(\zeta):=\frac{\zeta}{\Gamma(\sigma)}
\int_0^\infty dt\;\frac{t^{\sigma -1}e^{-t}}
{1-\zeta \epsilon e^{-t}},
\end{equation}
analytic in 
${\C}\setminus [1, \infty [$ (or ${\C}\setminus ]-\infty,-1]$ ) if 
$\epsilon=1$ (or $\epsilon=-1$). If $|\zeta|<1$, they are given 
by the following expansion:
$$f_\sigma^{\epsilon}(\zeta)=\sum_{n=1}^\infty\frac{\epsilon^{n-1}\zeta^n}
{n^\sigma}\;.$$

Now it is rather easy  to verify that for any $N\geq 0$, the 
multiple derivative 
$\partial_\omega^NP_\infty(\beta,\cdot,\omega_0)$ exists and defines 
an analytic function on 
$D_+$ (or $D_-$). The main result of \cite{3} established the pointwise 
convergence,
$$ \lim_{L\to \infty}\partial_\omega^NP_L(\beta,\cdot,\omega_0)=
\partial_\omega^NP_\infty(\beta,\cdot,\omega_0):=
\chi_\infty^N(\beta,\cdot,\omega_0);\quad |z|<1.$$

Remember that we want to apply the Vitali Convergence Theorem 
(see \cite{9} or \cite{3}). Therefore, in order to conclude that 
$\chi_L^N(\beta,z,\omega_0)$ converges 
uniformly to $\chi_\infty^N(\beta,z,\omega_0)$ for all $z\in K$, the only 
remaining point  is to get  the uniform boundedness w.r.t.  $L$. More precisely, we 
will prove: 
\begin{theorem}\label{th1.1}
For all $N\geq 1$, for all $\beta>0$ and for all $\omega>0$, 
\begin{equation}\label{partie2}
\sup_{L>1} 
\sup_{z\in K}\left|\chi_L^N(\beta,z,\omega)\right|\leq\,\cst(\beta,K,\omega,N).
\end{equation}
\end{theorem}

Then putting this together with the pointwise convergence result near 
$z=0$ of \cite{4}, 
the final conclusion would be:
\begin{equation}\label{partie2233}
\lim_{L\to \infty} 
\sup_{z\in K}\left|\chi_L^N(\beta,z,\omega_0)-\chi_\infty^N(\beta,z,\omega_0)
\right|=0.
\end{equation}
\begin{remark}
Having uniform convergence \eqref{partie2233} with respect to $z$ allows  us   to prove  existence of  the thermodynamic limit   for  canonical susceptibilities (see \cite {3,  6})
\end{remark}

This paper is devoted to the proof of Theorem \ref{th1.1}. Note that the 
theorem is an immediate consequence of the following estimate:
\begin{equation}\label{formule-principale}
\sup_{\xi\in \mathcal{C}_K}\sup_{z\in K}\left|\frac{\partial^N \tr\; g_L}
{\partial\omega^N}\,(\beta,\xi,z,\omega_0)\right|
\leq\,L^3\,\cst(\beta,K,N,\omega_0),
\end{equation}
which would imply via \eqref{deffkon} that the generalized 
susceptibilities are uniformly bounded in $L$.

\subsection{Strategy of the proof}

From now on, we omit the parameters $\xi$ and $z$ in the definition of 
$g_L$ in 
order to simplify notation. Fix  $\beta >0$ and $\omega_0\geq 0$.  Let 
$\Omega\subset\R$ be a compact interval containing $\omega_0$. If 
$\omega\in\Omega$, we denote by $\delta\omega:=\omega-\omega_0$. The main idea 
of the proof is to derive an equality of the following type:
\begin{equation}\label{zastrategi}
\tr\; g_L (\beta,\omega)=\tr\; g_L (\beta,\omega_0)+
\sum_{j=1}^N (\delta\omega)^j a_j(\beta,\omega_0)+(\delta\omega)^{N+1}
\mathcal{R}_L(\beta,\omega,N),
\end{equation}    
where the coefficients $a_j(\beta,\omega_0)$ grow at most like $L^3$ 
uniformly in $\xi$ and $z$, while 
the remainder $ \mathcal{R}_L(\beta, \cdot,N)$ is a smooth function near 
$\omega_0$. Then 
since we know that $\tr\; g_L (\beta,\cdot)$ is smooth, we must have 
$$  \frac{\partial^N\tr\; g_L}
{\partial\omega^N}\,(\beta,\omega_0)=N! a_N(\beta,\omega_0),$$
and this would finish the proof. In order to  achieve  this program, 
we will have to do two things.

First step: with the 
help of magnetic perturbation theory we will find a regularized 
expansion in $\delta\omega$ for $g_L$ of the form
\begin{equation}\label{zastrategi21}
g_L(\beta, \omega)=\sum_{n=0}^N (\delta\omega)^n 
g_{L,n}(\beta,\omega)+ R_{L,N}(\beta,\omega,N), 
\end{equation}
which holds in the sense of trace class operators, and the remainder has the 
property that $\frac{1}{(\delta\omega)^{N+1}}R_{L,N}(\beta,\omega)$ is smooth near
$\omega_0$ in the trace class topology. The 
operator-coefficients $g_{L,n}(\beta, \omega)$ {\it will still depend on} 
$\omega$, but 
in a more convenient way. That is, they are sums, products, or integrals of 
products of {\it regularized operators}, see 
\eqref{def-op-regularise}. 
This result is precisely stated in Theorem 
\ref{th3.4}.

Second step: show that for each $0\leq n\leq N$ we can write
\begin{equation}\label{zastrategi22}
\tr\; g_{L,n} (\beta,\omega)=\sum_{j=0}^N (\delta\omega)^j s_{L,j,n}(\beta,\omega_0)+
(\delta\omega)^{N+1}\mathcal{R}_{L,N}(\beta,\omega,N),
\end{equation} 
where the remainder $\mathcal{R}_{L,n}(\beta, \cdot)$ is smooth near $\omega_0$. Now 
the coefficients $s_{L,j,n}(\beta,\omega_0)$ are finally independent of 
$\omega$, and grow at most like $L^3$. This is done in the last section. 

Finally, if we combine \eqref{zastrategi22} with \eqref{zastrategi21}, we 
immediately obtain \eqref{zastrategi}. 

Now let us discuss why a more direct approach only based on trace norm 
estimates cannot work. 
Recall that the map $\omega\to W_L(\beta,\omega)\in B_1$ is real analytic, 
hence $\frac{\partial^N W_L}{\partial\omega^N}$ is well defined in 
$B_1(L^2(\Lambda))$, and we have the estimate (see 
\cite{1} and \cite{4}) : 
\begin{equation}\label{norme-I1-deriv-WL}
\left\|\frac{1}{N!}\,\frac{\partial^N W_L}{\partial\omega^N}(\beta,\omega_0)
\right\|_{B_1}\leq\,c_N\,
\frac{L^{3+N}\,(1+\beta)^{sN}}{\beta^{\frac{3}{2}}\left[\frac{N-1}{4}\right]!},
\end{equation}
where  $c_N$ is a positive constant which depends on $N$, $\omega_0$ and $s$. 
Now if we use the Leibniz rule of differentiation for the product 
which defines  the operator $g_L(\beta,\omega)$ (see \eqref{gebetatauom}), and estimate traces by 
trace norms we obtain: 
\begin{equation}
\left|\frac{\partial^N\tr\; g_L}{\partial\omega^N}(\beta,\omega_0)\right|
\leq\left\|\frac{\partial^N g_L}{\partial\omega^N}(\beta,\omega_0)
\right\|_{B_1}\leq   L^{3+N} {\cst}\,(\beta,K,\omega_0,N).
\end{equation}
Now this is definitely not good enough, and we have to find a more convenient 
expansion, as described in \eqref{zastrategi22}. This will be done in the 
next sections.

\section{Regularized expansion of $W_L$}%
\label{section2}

It has been shown in \cite{3} and \cite{4} that by using gauge 
invariance one can control the linear growth of the magnetic vector potential 
$\A$. The price one pays is the introduction of an antisymmetric phase factor, 
which disappears though when one takes the trace. Let us now  show how this 
works  for  the operator$W_L$. 

Fix $\omega_0\geq 0$ and $\beta >0$. Let  $\omega \in \C $ and $\delta\omega$ as 
above. Let us define the magnetic phase:
\begin{equation}
\phi(\x,\x'):=\x\cdot\A(\x')=\frac{1}{2}(x_2x_1'-x_1 x_2')=-\phi(\x',\x) ; \quad (\x,\x') \in \Lambda^2.
\end{equation}
If $T(\omega_0)$ is a bounded operator with
an integral kernel $t(\cdot,\cdot,\omega_0)$, 
then the notation $\widetilde{T}(\omega)$ will refer to
the regularized operator associated to $T(\omega_0)$ which has the kernel:
\begin{equation}\label{def-op-regularise}
\widetilde{t}(\x,\x',\omega):=e^{i \delta \omega \phi(\x,\x')}
t(\x,\x',\omega_0) ; \quad (\x,\x') \in \Lambda^2.
\end{equation}
We will very often use the Schur-Holmgren criterion of boundedness for 
integral operators (see \cite{HS}), which states that if $T$ has an integral kernel 
$t(\x,\x')$, then:
\begin{equation}\label{critere-r_LN-borne}
||T||\leq \left \{\sup_{\x'\in\Lambda}\int_{\Lambda}|t(\x,\x')|d\x\,
\cdot \sup_{\x\in\Lambda}\int_{\Lambda}|t(\x,\x')|d\x'\;\right \}^{1/2}.
\end{equation}

We denote by $G_L(\cdot,\cdot,\beta,\omega)$ the kernel of $W_L(\beta,\omega)$. 
We define two other bounded operators $R_{1,L}$ and $R_{2,L}$ by 
their kernels,
\begin{align}\nonumber
R_{1,L}(\x,\x',\beta) & :=  \A(\x-\x')\cdot \left [i\nabla_\x+
\omega_0\A\left(\x\right)\right ]G_L(\x,\x',\beta,\omega_0),\\ 
\label{def_R1L-R2L-RL-bis}  R_{2,L}(\x,\x',\beta) & := 
 \frac{\A^2(\x-\x')}{2}\,G_L(\x,\x',\beta,\omega_0); \quad (\x,\x') \in \Lambda^2.
\end{align}
Then consider the corresponding regularized operators $\widetilde{W}_L$,
$\widetilde{R}_{1,L}$, $\widetilde{R}_{2,L}$.
Let us state here two important estimates, the first one is just the 
diamagnetic inequality, while the second one was obtained in \cite{5},
\begin{align}\label{ineg-diam}
&\left|G_L(\x,\x',\beta,\omega_0)\right|  \leq 
  G_{\infty}(\x,\x',\beta,0)=\frac{1}{(2\pi \beta)^{3/2}}\exp \left ( -
\frac{|\x-\x'|^2}{2\beta}\right ),\\ \label{fancyesty}
& \left|\left [i\nabla_\x+\omega_0 \A(\x)\right]
     G_L(\x,\x',\beta,\omega_0)\right| \leq  C (1+\omega_0)^3\frac{(1+\beta)^5}
{\sqrt{\beta}}\, G_{\infty}(\x,\x',8\beta,0),
\end{align}
on $\Lambda^2$, where $C>0$ is a numerical constant.  A straightforward application of the 
Schur-Holmgren criterion gives us the 
following operator norm estimates
\begin{align}\label{fancyesty2}
\|\widetilde{W}_L(\beta,\omega)\|\leq 1,\quad 
\|\widetilde{R}_{i,L}(\beta,\omega)\|\leq
C_0\,(1+\omega_0)^3
(1+\beta)^5,
\end{align}
where $i=1,2$ and $C_0 >0$ is a numerical constant. 

For $i_1,...,i_n\in \{1,2\}$, define 
\begin{equation}\label{domeniuss}
D_n(\beta)=\{(\tau_1,...,\tau_n)\in \R^n\; :\; 0<\tau_n <...<\tau_1<\beta\}.
\end{equation}
Introduce the operator 
norm convergent Bochner integrals:
\begin{align}\nonumber
&  I_{n,L}(i_1,...,i_n)(\beta,\omega):=
\int_{D_n(\beta)}d\tau\,\widetilde{W}_L(\beta-\tau_1,\omega)
\widetilde{{R}}_{i_1,L}(\tau_1-\tau_2,\omega) \\ \label{InL2-def}
& \cdot\widetilde{{
    R}}_{i_2,L}(\tau_2-\tau_3,\omega)...\widetilde{{
    R}}_{i_{n-1},L}(\tau_{n-1}-\tau_n,\omega)\widetilde{{
    R}}_{i_n,L}(\tau_n,\omega),
\end{align}
and
\begin{align}\nonumber
&  J_{n,L}(\beta,\omega):=\int_{D_n(\beta)}d\tau\,W_L(\beta-\tau_1,\omega)
\widetilde{{R}}_{L}(\tau_1-\tau_2,\omega) \\ \label{JNLdef2}
& \cdot\widetilde{{R}}_{L}(\tau_2-\tau_3,\omega)...
\widetilde{{R}}_{L}(\tau_{n-1}-\tau_n,\omega)
\widetilde{{R}}_{L}(\tau_n,\omega).
\end{align}
Here we used the notation:
\begin{equation}\label{domeniuss2}
\widetilde{{
    R}}_{L}(\beta,\omega)=(\delta\omega)\widetilde{{
    R}}_{1,L}(\beta,\omega)+(\delta\omega)^2\widetilde{{
    R}}_{2,L}(\beta,\omega).
\end{equation}
  By \eqref{fancyesty2} the operators $I_{n,L}(i_1,...,i_n)$ and $J_{n,L}$ belong to 
$B(L^2(\Lambda))$. We will show below that in fact  their belong to $B_1(L^2(\Lambda))$ and their trace norm is of order $L^3$.
Denote by $\chi_n^j(i_1,...,i_n)$ the
characteristic function of the set 
$$\{(i_1,...,i_n)\in\{1,2\}^n\;:\;\sum_{k=1}^n i_k=j\}.$$

\begin{proposition}\label{th2.1}
Fix  $\omega_0>0$, and $N\geq 1$. Set $\delta \omega= \omega-\omega_0, \omega \in \C$. Then we have the following identity 
in $B_1(L^2(\Lambda))$:
\begin{equation}
W_L(\beta,\omega)=\widetilde{W}_L(\beta,\omega)+\sum_{n=1}^{N}
(\delta\omega)^{n}\,W_{L,n}(\beta,\omega)+
R_{L,N}^{(1)}(\beta,\omega), \label{pseudoWL}
\end{equation}
where $\omega \to \frac{1}{(\delta\omega)^{N+1}}R_{L,N}^{(1)}(\beta,\omega)$  is a
smooth $\ B_1$ operator valued function in  $\omega$. 

The coefficients of the above expansion are given by:
\begin{align}\label{def-WLN}
W_{L,n}(\beta,\omega):=\sum_{k=1}^n(-1)^k
\sum_{i_j\in\{1,2\}}\chi_{k}^{n}(i_1,...,i_k)I_{k,L}(i_1,...,i_k)(\beta,\omega),
\end{align}
and the reminder reads as:
\begin{align}\label{emmerdement}
&R_{L,N}^{(1)}(\beta,\omega):=
\sum_{n=N+1}^{2N}(\delta\omega)^n\sum_{k=1}^{N}(-1)^k
\sum_{i_j\in\{1,2\}}\chi_k^n(i_1,...,i_k) \\ \nonumber
& \cdot I_{k,L}(i_1,...,i_k)(\beta,\omega)+(-1)^{N+1}J_{N+1,L}(\beta,\omega).
\end{align}
\end{proposition}

\noindent {\bf Proof.} 
This proposition was proved in \cite{4} in the sense of bounded operators. 
Again by using Remark 3.4 in \cite{4}, the operator 
$\widetilde{W}_L(\beta,\omega)$ as well as $W_{L,n} (\beta,\omega)$ belong to 
$B_1(L^2(\Lambda))$. The same argument holds for the remainder 
$R_{L,N}^{(1)}(\beta,\omega)$. The fact that the operator 
$\omega \to \frac{1}{(\delta\omega)^{N+1}}R_{L,N}^{(1)}(\beta,\omega)$ is a smooth $B_1$ operator
valued function  near $\omega_0$ follows from definitions \eqref{emmerdement}, 
\eqref{InL2-def} and \eqref{JNLdef2}. \qed

\begin{remark}\label{remarque333}
It is important to note that 
the coefficients $W_{L,N}$ in (\ref{def-WLN}) still depend on $\omega$, 
but only through the magnetic phases. Let $\omega \in\R$.
From the estimates \eqref{fancyesty2} and the definitions in \eqref{InL2-def} 
and  \eqref{JNLdef2}, and after integration over the $\tau$ variables, clearly we 
get as bounded operators:
\begin{equation}\label{briet1111}
||R_{L,N}^{(1)}(\beta,\omega)||\leq {\rm const} \; |\delta\omega|^{N+1},
\end{equation}
where the constant is uniform in $L>1$. 

\end{remark}

\section{Regularized expansion of $g_L$}%

We will now try to get  a similar expansion for $g_L$, aiming to obtain 
\eqref{zastrategi21}.  Fix $\beta >0$  and $\omega_0 >0$. The parameters
 $z$ and $\xi$ which enter the definition of $g_L$ are fixed as in 
(\ref{gebetatauom}), and the estimates we make must be uniform w.r.t. them. Here 
$\omega$ is real and $\delta \omega$ is as above.
It was shown in \cite{5} that:
\begin{lemma}\label{lemaexploc} Let $\omega  \in\R$.
The trace class
operator $g_L$  admits  a continuous integral kernel. Moreover, there exist 
two positive constants $C$ and $\alpha$, both independent of $L$, such that the
integral kernel satisfies
\begin{equation}\label{estim-kernel-gL}
|g_L(\x,\x'; \omega)|\leq C\; e^{-\alpha |\x-\x'|}.
\end{equation}
\end{lemma}
Looking at \eqref{gebetatauom}, we see that we need a regularized expansion for 
the operator $(\xi-z W_L(\beta,\omega))^{-1}$. 

Using \eqref{briet1111} with $N=1$ we get for $\omega \in \R$,
\begin{equation}\label{difference-WL-WLtilde}
\|W_L(\beta,\omega)-\widetilde{W}_L(\beta,\omega)\|\leq { C_1} \;
|\delta \omega|,
\end{equation}
where C$_1$ is a  $L$   independent constant . Hence  choose
$|\delta\omega|$ is small enough such that
\begin{equation}\label{boubou1}
{C_1} M |\delta \omega|\; \;(\sup_{z\in K}|z|)\;  < 1/2,
\end{equation}
$C_1$   and $M$     being   respectively   defined  in \eqref{boubou1}  
and  \eqref{uniffbou}. Then the operator $1-z(\widetilde{W}_L-W_L)(\xi-zW_L)^{-1}$ is invertible and 
its inverse has a norm less than $2$. Hence by choosing $\Omega$ to be a small 
enough interval around $\omega_0$ we get
\begin{equation}\label{boubou2}
\sup_{L>1}\sup_{\omega\in\Omega}\sup_{\xi\in \mathcal{C}_K }\sup_{z\in K}
\left \Vert [\xi-z
 \widetilde{W}_L(\beta,\omega)]^{-1} \right \Vert \leq 2M.
\end{equation}
Then we can write:
 \begin{align}\label{boubou10}
& (\xi-z W_L(\beta,\omega))^{-1}\\ \nonumber
& =(\xi-z
\widetilde{W}_L(\beta,\omega))^{-1}\sum_{n=0}^\infty z^n 
\{[W_L(\beta,\omega)-\widetilde{W}_L(\beta,\omega)](\xi-z
\widetilde{W}_L(\beta,\omega))^{-1}\}^n,
\end{align}
and thus we reduced the problem to the study of 
$(\xi-z\widetilde{W}_L(\beta,\omega))^{-1}$. 

In order to  get a convenient expansion for this inverse, we need to  
introduce some new notation. Let $\x,\x' \in \Lambda$. Define for $N\geq 1$:
\begin{align}\label{def-kernel-rLn}
r_{L,N}(\x,\x';\beta):=
-z\int_{\Lambda} \frac{[i\,{\rm
    fl}(\x,\y,\x')]^{N}}{N!}\,G_L(\x,\y;\omega_0)
g_L(\y,\x';\omega_0)d\y,
\end{align}
where 
\begin{equation}\label{flucsibucsi}
{\rm fl}(\x,\y,\x'):=\phi(\x,\y)+\phi(\y,\x')+\phi(\x',\x)=\frac{1}{2}
{\bf e}_3\cdot [(\y-\x)\wedge (\y-\x')]; 
\end{equation}
with $ {\bf e}_3=(0,0,1)$, denotes the magnetic flux through the triangle defined by $\x$, $\y$, 
 and  $\x'$. Similarly let $\omega \in \R$ and define the bounded operator $r_L(\beta,\omega)$ 
 and $\hat r_L(\beta,\omega)$ whose kernel is given by:
\begin{align} \nonumber 
r_L(\x,\x'; \beta,\omega)&=-z\int_{\Lambda}\,
\left(e^{i\delta\omega{\rm fl}(\x,\y,\x')}-1\right)
G_L(\x,\y;\beta,\omega_0)g_L(\y,\x';\beta, \omega_0)d\y, \\
\hat r_L(\x,\x'; \beta,\omega)&=  e^{i\delta \omega \phi(\x,\x')}r_L(\x,\x'; \beta,\omega)
\label{boubou26}\end{align}
Notice that $\hat r_L$  does not coincide with the regularization  of $ r_L$
given by  \eqref {def-op-regularise}.
The operator  $r_L$  and $\hat r_L$  are  related to the
operators $r_{L,N}$  and  $\tilde r_{L,N}$ respectively  by
\begin{equation}\label{rL=serie}
r_L(\beta,\omega)=\sum_{k=1}^\infty (\delta\omega)^k
r_{L,k}(\beta);\quad   \hat r_L(\beta,\omega)=\sum_{k=1}^\infty (\delta\omega)^k
\tilde r_{L,k}(\beta).
\end{equation}
Note that  by using  the Schur Holmgren criterion, the Lemma \ref{lemaexploc}, \eqref{ineg-diam} and the fact that  ${\rm fl}(\x,\y,\x')$ is bounded  from above by $ L^2$ on $\Lambda$, we have the estimate:
\begin{equation}\label{rL=serie2}
\left \Vert r_L-\sum_{k=1}^N (\delta\omega)^k
r_{L,k}=\sum_{k=N+1}^\infty (\delta\omega)^k
r_{L,k}\right \Vert \leq \cst\; e^{L^2} |\delta\omega|^{N+1},
\end{equation}
 for some numerical positive  constant. The same estimate again holds true  for
$ \hat r_L(\beta,\omega)$ and  the corresponding  series   given in  \eqref{rL=serie}.

Let us now give some more precise estimates on the norms of these operators.
\begin{proposition}\label{th3.1}
Fix $N\geq 1$. There exist a positive constants $C_2$  
independent of $L>1$ such that for all $\omega \in \R$,
\begin{equation}\label{boubou4}
\max\left \{\max_{k=1}^N ||r_{L,k}(\beta)||,\; |\delta\omega|^{-1}
||r_L(\beta,\omega)||\right \}\leq C_2,
\end{equation}
and 
 \begin{equation}\label{boubou5}
\max\left \{\max_{k=1}^N ||r_{L,k}(\beta)||_{B_2},\; |\delta\omega|^{-1} 
||r_L(\beta, \omega)||_{B_2}\right 
\}=C_2 \cdot  L^{3/2}.
\end{equation}
These estimates also hold true for the regularized operators in the sense of 
\eqref{def-op-regularise} and $ \hat r_L(\beta,\omega)$.
\end{proposition}
\noindent {\bf Proof.} First, note that 
$|{\rm fl}(\x,\y,\x')|\leq |\x-\y|\,|\y-\x'|$, see \eqref{flucsibucsi}. 
The kernels present in the 
$\y$ integral are localized near their diagonal, see (\ref{ineg-diam}) and 
(\ref{estim-kernel-gL}). By 
extending the integral with respect to $\y$ over the whole $\R^3$, then using 
a fraction of the exponential decay in order to bound the polynomial growth from 
the flux, we obtain a constant independent of $L$ such that
\begin{equation}\label{estim-rLN-kernel}
|r_{L,k}(\x,\x'; \beta)|\leq\,C_3\cdot e^{-\frac{\alpha}{4}|\x-\x'|};\quad 
 (\x,\x' ) \in \Lambda^2, \quad 1\leq k\leq N.
\end{equation}
The same estimate holds for $r_L$. 
Now we can apply the Schur-Holmgren criterion (\ref{critere-r_LN-borne}) and get 
\eqref{boubou4}. The Hilbert-Schmidt estimates is also straightforward. 

\qed

The next proposition gives the necessary expansion of 
$(\xi-z \widetilde{W_L}(\beta,\omega))^{-1}$.
\begin{proposition}\label{th3.2} 
Fix $N\geq 1$ and $\omega_0>0$. Then if $|\delta\omega|$ is small enough, the 
following identity holds in $B(L^2(\Lambda))$:
\begin{align}\label{dev-resolvante-reg}
& (\xi-z
\widetilde{W}_L(\beta,\omega))^{-1}=\xi^{-1}
\left(1+\widetilde{g}_L(\beta,\omega)\right)\; 
[1+\xi^{-1}\hat{r}_L(\beta,\omega)]^{-1}\\ 
&=\xi^{-1}
\left(1+\widetilde{g}_L(\beta,\omega)\right) +
\sum_{n=1}^{N} (\delta\omega)^{n}\,S_{L,n}(\beta,\omega)+
R_{L,N}^{(2)}(\beta,\omega),\nonumber 
\end{align}
where $S_{L,N}$ is given by
\begin{align}\nonumber
& S_{L,N}(\beta,\omega):=\xi^{-1}\sum_{n=1}^{N}(-\xi^{-1})^{n}
\sum_{(i_1,...,i_n)\in {(N^{*})}^{n}}\chi_{n}^{N}(i_1,...,i_n)
\left(1+\widetilde{g}_L(\beta,\omega)\right)\\ \label{def-SLN-def}
& \cdot
\widetilde{r}_{L,i_1}(\beta,\omega)...
\widetilde{r}_{L,i_n}(\beta,\omega)\,
\end{align}
where the  remainder $R_{L,N}^{(2)}(\beta,\omega)$  has the property that 
the bounded operator valued function $\omega \to (\delta\omega)^{-N}R_{L,N}^{(2)}(\beta,\omega)$ is smooth around $\omega_0$ 
and moreover,  there exists a constant (possibly)  depending  on $L$ such that
\begin{equation}\label{estim-reste-2}
\|R_{L,N}^{(2)}(\beta,\omega)\|\leq\,\cst \; |\delta\omega|^{N+1}.
\end{equation}
\end{proposition}
\noindent {\bf Proof.} We start with the following resolvent equation,
\begin{equation}\label{eq-resolvantes}
(\xi-z \widetilde{W}_L(\beta,\omega))^{-1}=\xi^{-1}+(\xi-z
\widetilde{W}_L(\beta,\omega))^{-1} z \widetilde{W}_L(\beta,\omega)\xi^{-1}.
\end{equation}
Now the next identity 
is  very important, and it is obtained by  a straightforward  calculation from 
\eqref{flucsibucsi} and the definition of $g_L$ (see also Proposition 13 in \cite{5})
 \begin{equation} \label{boubou7}
[\xi-z\widetilde{W}_L(\beta,\omega)]\widetilde{g}_L(\beta,\omega)=
z\widetilde{W}_L(\beta,\omega) +\hat{r}_L(\beta,\omega).
\end{equation}
If one multiplies with an inverse both sides of the above equality we get:
\begin{equation}\label{lien-avec-gtildeL}
(\xi-z\widetilde{W}_L(\beta,\omega))^{-1}z\widetilde{W}_L(\beta,\omega)
=\widetilde{g}_L(\beta,\omega)-(\xi-z \widetilde{W}_L(\beta,\omega))^{-1}
\hat{r}_L(\beta,\omega).
\end{equation}
We know from Proposition \ref{th3.1} that we can find a constant $C_2$ independent of 
$L$ such that
\begin{equation}\label{norm-rL-tilde}
\| {\hat r}_L( \beta,\omega)\|\leq C_2\; |\delta\omega|.
\end{equation}
Let us use (\ref{lien-avec-gtildeL}) in (\ref{eq-resolvantes}), and isolate the 
inverse we are interested in
\begin{equation}\label{boubou9}
[\xi-z \widetilde{W}_L(\beta,\omega)]^{-1}[1+\xi^{-1}\hat{r}_L(\beta,	\omega)]
=\xi^{-1}[1+\widetilde{g}_L(\beta,\omega)].
\end{equation}
Now if $|\delta\omega|$ is small enough, $1+\xi^{-1} {\hat r}_L(\omega)$ 
is invertible and \eqref{dev-resolvante-reg} follows. Moreover, expressing the 
inverse by a finite Neumann-type expansion, 
$$[1+\xi^{-1}\hat{r}_L]^{-1}=
 \sum_{k=0}^{N} (-\xi)^{-k}{\hat r}_L^k + 
[1+\xi^{-1}\hat{r}_L]^{-1}(-\xi)^{-(N+1)}\hat{r}_L^{(N+1)},$$
and using \eqref{rL=serie} we can identify the operators 
$S_{L,N}(\beta,\omega)$ as given in \eqref{def-SLN-def}, while the reminder reads 
as
\begin{align}
&R_{L,N}^{(2)}(\beta,\omega):= (-\xi)^{-(N+1)}[\xi-z \widetilde{W}_L(\beta,\omega)]^{-1}
\hat{ r}_L^{N+1}
(\beta,\omega)+ \xi^{-1}[1+\widetilde{g}_L(\beta,\omega)]\nonumber 
\\
&\cdot\sum_{k=N+1}^{\infty}\,(\delta\omega)^{k}\sum_{n=1}^{N}(-\xi^{-1})^{n}
\sum_{(i_1,...,i_n)\in {(N^{*})}^{n}}\chi_{n}^{k}(i_1,...,i_n){\widetilde r}_{L,i_1}
(\beta,\omega)...\widetilde{r}_{L,i_n}(\beta,\omega).
\end{align}
Let us now identify the term in $(\delta\omega)^{N+1}$ which appears 
in the estimate (\ref{estim-reste-2}). For the first term of the remainder it 
comes from (\ref{norm-rL-tilde}), while for the second one 
it comes from the fact that
the series begin with the index $N+1$ (see \eqref{rL=serie2}). \qed

We are now ready to give a convenient expansion for the operator 
$[\xi-z W_L(\beta,\omega)]^{-1}$. First  we need 
some new notation. We introduce the following operators,
\begin{align}
&S_{L,0}(\beta,\omega):=\xi^{-1}[1+\widetilde{g}_L(\beta,\omega)],
\label{eselzero} \\
&T_{L,N}(\beta,\omega):=\sum_{n=1}^{N}z^{n}\sum_{0\leq i_k\leq
  N, 1\leq j_k\leq N}\chi_{2n+1}^{N}(i_0,j_1,i_1,...,j_n,i_n)S_{L,i_0}(\beta,\omega)
\nonumber\\ \label{def-TLN-def} 
&\cdot W_{L,j_1}(\beta,\omega)S_{L,i_1}(\beta,\omega)\,...\,
W_{L,j_n}(\beta,\omega)S_{L,i_n}(\beta,\omega),\,\,\,N\geq 1.
\end{align}
Since the operators $W_{L,j}$ and $S_{L,i}$ defined in Propositions 
\ref{th2.1} and \ref{th3.2} are uniformly bounded in $L$, this is also true for 
$T_{L,N}$. 
\begin{corollary}\label{th3.3} 
Fix $N\geq 1$ and $\omega_0\geq 0$. If $\vert \delta\omega \vert $ is small enough, then the 
following identity holds in $B(L^2(\Lambda))$:
\begin{align}\label{dev-res}
[\xi-z W_L(\beta,\omega)]^{-1}=[\xi-z \widetilde{W}_L(\beta,\omega)]^{-1}
 +\sum_{n=1}^{N}
(\delta\omega)^{n}\,T_{L,n}(\beta,\omega)
+R_{L,N}^{(3)}(\beta,\omega),
\end{align}
where the remainder $R_{L,N}^{(3)}(\beta,\omega)$ has the property 
that the bounded  operator valued function
 $\omega \to \frac{1}{(\delta\omega)^N} R_{L,N}^{(3)}(\beta,\omega)$ is smooth near 
$\omega_0$, and there exists a constant  (possibly) depending on $L$ such that:
\begin{equation}\label{estim-reste-3}
\|R_{L,N}^{(3)}(\beta,\omega)\|\leq \,{\cst}\,|\delta\omega|^{N+1} .
\end{equation}
\end{corollary}
\noindent {\bf Proof.} The result follows after inserting 
the estimates from the previous proposition into formula
\eqref{boubou10}, having used the notation introduced in \eqref{eselzero}, 
\eqref{def-TLN-def}, \eqref{pseudoWL} and \eqref{def-WLN}. The rest is just 
a tedious bookkeeping of various terms.\qed

We finally are in the position of writing ''the right'' expansion for the 
operator  $g_L(\beta,\omega)$ as announced in \eqref{zastrategi21}.
\begin{theorem}\label{th3.4} 
Fix $N\geq 1$ and $\omega_0\geq 0$. If $ \vert \delta\omega \vert$ is small enough, then 
the following equality takes place in $B_1(L^2(\Lambda))$:
\begin{equation}\label{dev-g-L-bornes}
g_L(\beta,\omega)=g_{L,0}(\beta,\omega)+
\sum_{n=1}^{N} (\delta\omega)^{n}\,g_{L,n}(\beta,\omega)+
R_{L,N}^{(4)}(\beta,\omega),
\end{equation}
where 
\begin{equation}\label{maisonfada1}
g_{L,0}(\beta,\omega):=[\xi-z
\widetilde{W}_L(\beta,\omega)]^{-1}z\widetilde{W}_L(\beta,\omega).
\end{equation}  
and $g_{L,n}$ are given by ($N\geq 1$),
\begin{align} g_{L,N}(\beta,\omega):=\sum_{n=1}^{N}
\left[S_{L,N-n}(\beta,\omega)
zW_{L,n}(\beta,\omega)
\right. \label{def-coef-gLn-borne} \left.+
T_{L,n}(\beta,\omega)zW_{L,N-n}(\beta,\omega)\right],
\end{align}
where $W_{L,0}:=\widetilde{W}_L$ and the remainder $\frac{1}{(\delta\omega)^N} R_{L,N}^{(4)}(\beta,\omega)$ has the property that  the $B_1$ operator valued function $\omega \to
\frac{1}{(\delta\omega)^N} R_{L,N}^{(4)}(\beta,\omega)$
 is smooth near 
$\omega_0$ and there exists a  positive constant (possibly)
depending on $L$ such that:
\begin{equation}\label{estim-reste-7}
\|R_{L,N}^{(4)}(\beta,\omega)\|_{B_1}\leq \,{\cst}\,|\delta\omega|^{N+1} L^3.
\end{equation}
\end{theorem}
\noindent {\bf Proof.} First we multiply the $B_1(L^2(\Lambda))$ 
expansion \eqref{pseudoWL} of the 
semigroup  with
the expansion \eqref{dev-res}  of the resolvent valid in $B(L^2(\Lambda))$. 
Thus one obtains in $B_1(L^2(\Lambda))$,
\begin{align}\nonumber
&  g_L(\beta,\omega)=[\xi-z
\widetilde{W}_L(\beta,\omega)]^{-1}z\left(\widetilde{W}_L(\beta,\omega)+
\sum_{n=1}^{N}(\delta\omega)^{n}W_{L,n} (\beta,\omega)\right)\\ \nonumber
& +\sum_{n=1}^{N}(\delta\omega)^{n}\,T_{L,n}(\beta,\omega)z 
{\widetilde W_L}(\beta,\omega) +
\sum_{n=1}^{N}\sum_{k=1}^{N}(\delta\omega)^{n+k}\,T_{L,n}(\beta,\omega)
zW_{L,k}(\beta,\omega) \\ \nonumber
& +[\xi-z W_L(\beta,\omega)]^{-1}z
R_{L,N}^{(1)}(\beta,\omega)+ 
R_{L,N}^{(3)}(\beta,\omega) z\left(\sum_{n=1}^{N}(\delta\omega)^{n}\,
W_{L,n}(\beta,\omega)\right)\\ \label{dev-gL-non-reg}
&  +R_{L,N}^{(3)}(\beta,\omega) z \widetilde{W}_L(\beta,\omega).
\end{align}
The last two lines will  give  a remainder $R_{L,N}^{(4_a)}(\beta,\omega)$, 
whose properties can be read out of those of the previous ones. This 
remainder has the same properties as $R_{L,N}^{(4)}(\beta,\omega)$, and in fact 
it is a part of it. The rest of the proof is just algebra, and amounts to 
identify the right factors which enter the definition of $g_{L,n}(\beta,\omega)$ 
and the expression of the full  remainder. Here one must use 
\eqref{dev-resolvante-reg} and the notation introduced in \eqref{def-SLN-def}, 
 \eqref{eselzero} and \eqref{def-TLN-def}. The proof is over.\qed

\section{Expansion  of the trace of $g_L$. The uniform bound.}

We have almost all ingredients needed for proving \eqref{zastrategi22}. 
Let $\beta >0$, $\omega_0 >0$ and $\omega \in \Omega $ as in the Section  1.1. We now 
need to take the trace in \eqref{dev-g-L-bornes}. Let us begin with 
the trace of the operator $g_{L,0}(\beta,\omega)$ (see 
\eqref{maisonfada1}). 
If we use \eqref{lien-avec-gtildeL}, and then  
\eqref{dev-resolvante-reg} , \eqref{eq-resolvantes} and \eqref{eselzero}, we can write:
\begin{align}\label{maisonfada2}
&g_{L,0}(\beta,\omega)=\widetilde{g}_L(\beta,\omega)-
\xi^{-1}\hat{r}_L(\beta,\omega)-\xi^{-1}[\xi-z
\widetilde{W}_L(\beta,\omega)]^{-1}z\widetilde{W}_L(\beta,\omega)
{\hat r}_L(\beta,\omega)\nonumber \\
&=\{\widetilde{g}_L(\beta,\omega)-
\xi^{-1} {\hat r}_L(\beta,\omega)\}-\xi^{-1}z\sum_{k=0}^N
(\delta\omega)^k S_{L,k}(\beta,\omega)\widetilde{W}_L(\beta,\omega)
{\hat r_L}(\beta,\omega)\nonumber \\
&-\xi^{-1}zR_{L,N}^{(2)}(\beta,\omega)\widetilde{W}_L(\beta,\omega)
\hat{r}_L(\beta,\omega).
\end{align} 
Apriori, this identity only 
holds  in the  bounded operators sense. But we know that 
$\widetilde{W}_L(\beta,\omega)$ is a trace class operator. It means that 
the operator
\begin{align}\label{maisonfada3}
M(\beta,\omega):=\widetilde{g}_L(\beta,\omega)-\xi^{-1}{\hat r}_L(\beta,\omega)
\end{align} 
is a trace class operator, since all other operators in 
\eqref{maisonfada2} are trace class. Note that the two individual terms in 
$M(\beta,\omega)$ might not be trace class. 
Now since $M(\beta, \omega)$ has a continuous integral kernel $M(\cdot,\cdot;\beta,\omega)$ 
(see Lemma \ref{lemaexploc} and \eqref{boubou26}), 
its trace will be given by:   
\begin{align}\label{maisonfada4}
\tr \; M(\beta,\omega)& =\int_\Lambda M(\x,\x;\beta,\omega)d\x \\ 
&=
\int_\Lambda\widetilde{g}_L(\x,\x;\beta,\omega)d\x-\xi^{-1}\int_\Lambda
\hat{r}_L(\x,\x;\beta;\omega)d\x\nonumber \\
&=\int_\Lambda g_L(\x,\x;\beta,\omega_0)d\x-\xi^{-1}\int_\Lambda 
r_L(\x,\x;\beta,\omega)d\x. \nonumber
\end{align} 
The last line is very important, since it shows that the ''tilde'' 
disappears when we take the trace. This is because the magnetic phase 
$\phi(\x,\x)=0$ for all $\x$. But now 
$\int_\Lambda g_L(\x,\x;\beta,\omega_0)d\x =\tr \; g_L(\beta,\omega_0)$, and 
we here recognize the very first term on the right hand side of 
\eqref{zastrategi}. Now if we use \eqref{def-kernel-rLn} 
\eqref{boubou26} and  \eqref{rL=serie} we can write:
\begin{align}\label{maisonfada9}
\tr \; M(\beta,\omega)& =\tr \; g_L(\beta,\omega_0)-\xi^{-1}
\sum_{n=1}^N(\delta\omega)^n\int_\Lambda r_{L,n}(\x,\x; \beta,\omega_0)d\x 
\nonumber \\
&+(\delta\omega)^{N+1}\mathcal{R}_L^{(1)}(\beta,\omega,N),
\end{align}
where $\omega \to \mathcal{R}_L^{(1)}(\beta,\omega,N)$ is a  smooth function in $\omega$ near 
$\omega_0$. Moreover, due to \eqref{estim-rLN-kernel} we obtain that the above 
integrals grow at most like $L^3$, as required.

But there are several other terms which remain to be considered in 
\eqref{maisonfada2} and \eqref{dev-g-L-bornes}. They are respectively 
$\tr\; \{S_{L,k}(\beta,\omega)\widetilde{W}_L(\beta,\omega)
\widetilde{r}_L(\omega)\}$ and $\tr\; g_{L,n}(\beta,\omega)$.

These traces have two important things in common. First, we always take 
the trace of a product of integral operators with continuous kernels. Second, 
they all still depend on $\delta\omega$, but only through the magnetic phases; 
all factors are regularized operators, as defined in 
\eqref{def-op-regularise}. We will now try to discuss all these different 
terms in a unified manner.

Fix $\omega_0>0$. Let $\omega \in \R$ and $\delta \omega$ as above.
 Consider a product of operators of the form
$$T(\omega):=\widetilde{T}_0(\omega)\widetilde{T}_1(\omega)...
\widetilde{T}_n(\omega)$$
where $\widetilde{T}_i(\omega)$ are the regularized operators 
associated to some integral operators $T_i(\omega_0)$, $i=0,...,n$ (see \eqref{def-op-regularise}) and assume 
that this  product is of trace class. Denote by $t_i(\cdot,\cdot)$ 
the kernel of $T_i(\omega_0)$, which is supposed to be jointly continuous in 
$\x$ and $\x'$. We denote by ${\fl}_n$ the following flux related quantity
\begin{align}\nonumber
&{\fl}_1(\x,\y_1)=0,\quad {\fl}_n(\x,\y_1,...,\y_n)= 
\sum_{k=1}^{n-1}\fl(\x,\y_k,\y_{k+1})\\ \label{def-fln-fln}
& =\phi(\x,\y_1)+\phi(\y_1,\y_2)+...+\phi(\y_{n-1},\y_n)+\phi(\y_n,\x),\; 
n\geq 2.
\end{align}
Another important property of these operators is that their kernels are 
exponentially localized near the diagonal (see (\ref{ineg-diam}), 
(\ref{estim-kernel-gL}) and (\ref{estim-rLN-kernel})). Therefore 
 there exist two positive constants $C$ and $\alpha$, 
independent of $L$, such that:
\begin{align}\label{expobou}
\max_{i=0}^n
\left|t_i(\x,\x')\right|\leq\,C\,e^{-\alpha|\x-\x'|},\quad (\x,\x') \in\Lambda^2.
\end{align}

Then the diagonal value of the kernel of $T(\omega)$ reads as
\begin{align}\nonumber
& T(\x,\x,\omega)=\int_{\Lambda}d\y_1...\int_{\Lambda}d\y_n\,e^{i \delta\omega
\fl_{n}(\x,\y_1,...,\y_n)}t_0(\x,\y_1,\omega_0)t_1(\y_1,\y_2,\omega_0)...
\\ \label{kernelT}
& t_{n-1}(\y_{n-1},\y_{n},\omega_0)t_n(\y_{n},\x,\omega_0),
\end{align}
where we added together all individual phases from each regularized factor. 
Because we assumed that $T$ is a  trace class operator,  the trace of $T$ is
\begin{equation}\label{traceT}
\tr\,T (\omega)=\int_{\Lambda}T(\x,\x,\omega)d\x.
\end{equation}
For $m\geq 0$, $n\geq 1$, let us introduce the notation:
\begin{align}\nonumber
&d_{m,n}(L):=\int_{\Lambda}d\x\int_{\Lambda}d\y_1...\int_{\Lambda}
d\y_{n}[i\fl_{n}(\x,\y_1,..., \y_{n})]^m\, t_0(\x,\y_1,\omega_0)\\ \label{ii}
&t_1(\y_1,\y_2,\omega_0)...t_{n-1}(\y_{n-1},\y_{n},\omega_0)
t_n(\y_{n},\x,\omega_0).
\end{align}

\begin{lemma}\label{lemma11}
For every $m\geq 0$ and $n\geq 1$, there exists a constant 
independent of $L$  but depending on $m,n$  such that 
\begin{align}\label{boubou44}
|d_{m,n}(L)|\leq {\cst} L^3\,.
\end{align}
Moreover, for a given $N\geq 1$ we have
\begin{align}\label{boubou45}
\tr\; T(\omega)=\sum_{m=0}^N(\delta\omega)^m d_{m,n}(L)+
(\delta\omega)^{N+1}\mathcal{R}_L(\omega,N),
\end{align}
where  $ \omega \to \mathcal{R}_L(\omega,N)$ is a smooth  function near $\omega_0$.
\end{lemma}

\noindent {\bf Proof.} The equality \eqref{boubou45} comes straight out of 
\eqref{kernelT} and \eqref{traceT}. 

Now let us prove the estimate 
\eqref{boubou44}. We recall the following estimate \eqref{flucsibucsi}  
on the magnetic flux,
\begin{equation}\label{majfl}
\left|{\rm fl}(\x,\y,\z)\right|\leq |\x-\y| |\y-\z|.
\end{equation}
Then by induction  one has for all $n\geq 1$, 
\begin{align}
|\fl_n(\x,\y_1,...,\y_n)|\leq \left(|\x-\y_1|+|\y_1-\y_2|+...+
|\y_{n-1}-\y_n|\right)^2.
\end{align}
Therefore the polynomial growth induced by this flux is diagonalized, i.e. it 
only depends on differences between the  variables $y_i, y_{i+1}$. But due to \eqref{expobou}, we can 
write
\begin{align}\nonumber
&|d_{m,n}(L)|\leq \cst
\int_{\Lambda}d\x\int_{\Lambda}d\y_1...\int_{\Lambda}
d\y_{n} e^{-\frac{\alpha}{2}|\x-\y_1|}\\ \label{ii99}
&e^{-\frac{\alpha}{2}|\y_1-\y_2|}...e^{-\frac{\alpha}{2}|\y_{n-1}-\y_n|}
e^{-\frac{\alpha}{2}|\y_n-\x|},
\end{align}
 for some $L$ independent constant.  Here we used   the exponential decay to bound the polynomial factors and  
then we extend the $\y$ integrals to the whole $\R^3$. So   the volume 
growth is only given by the  integral over $\x$ in the r.h.s of  \eqref{ii99}. The proof the lemma is over.\qed

\subsection{Proof of \eqref{zastrategi} and of Theorem \ref{th1.1}}

We can now put together the results of this section and prove the key 
estimate \eqref{zastrategi}.  In Theorem \ref{th3.4} we obtained an 
expansion for $g_L(\beta,\omega)$ as announced in \eqref{zastrategi21}. When 
we take the trace of $g_L(\beta,\omega)$, the term 
$R_{L,N}^{(4)}(\beta,\omega)$ will only give a contribution to the remainder 
in \eqref{zastrategi}, hence we ignore it. 

Then the term $g_{L,0}(\beta,\omega)$ given in \eqref{maisonfada1} 
can be written as a sum between an operator $M(\beta,\omega)$ from 
\eqref{maisonfada3}, and a sum of 
operators of the type treated in Lemma \ref{lemma11}. Then from 
\eqref{maisonfada9} and the above mentioned lemma we can conclude that 
$(\partial_\omega^N \tr\;g_{L,0})(\beta,\omega_0)$ grows at most 
 like $L^3$. 

Finally,  looking at the contribution coming from $g_{L,n}(\beta,\omega)$, with $n\geq 1$. 
Using the same lemma, we obtain in a similar way that 
$(\partial_\omega^N \tr\;g_{L,n})(\beta,\omega_0)$ grows at most like the 
volume. We therefore conclude that 
$(\partial_\omega^N \tr\;g_{L})(\beta,\omega_0)$ behaves like $L^3$, 
uniformly in $\xi$ and $z$, and the proof of \eqref{formule-principale} 
is done.

\medskip
\noindent {\bf Acknowledgments.} The authors thank V. A. Zagrebnov, G.
Nenciu and N. Angelescu for many fruitful discussions.  H.C. acknowledges support from the Danish 
F.N.U. grant {\it  Mathematical Physics and Partial Differential
  Equations}.


\end{document}